\documentclass[preprint,showpacs,preprintnumbers,amsmath,amssymb]{revtex4}

\usepackage{graphicx}
\usepackage{dcolumn}
\usepackage{bm}

\begin{document}

\preprint{APS/123-QED}

\title{Modification of triaxial deformation and change of spectrum in $^{25}_{\ \Lambda}$Mg caused by $\Lambda$ hyperon}

\author{Masahiro Isaka and Hiroaki Homma}
\affiliation{%
Department of Cosmosciences, Graduate School of Science, Hokkaido University, Sapporo 060-0810, Japan
}%

\author{Masaaki Kimura}
\affiliation{%
Creative Research Institution (CRIS), Hokkaido University, Sapporo 001-0021, Japan
}%

\author{Akinobu Dot\'{e}}
\affiliation{
Institute of Particle and Nuclear Studies, KEK, Tsukuba, Ibaraki 305-0801, Japan
}%

\author{Akira Ohnishi}
\affiliation{
Yukawa Institute for Theoretical Physics, Kyoto University,Kyoto 606-8502, Japan
}%

%

\date{\today}

\begin{abstract}
The positive-parity states of $^{25}_{\ \Lambda}$Mg with a $\Lambda$ hyperon in $s$ orbit were studied with the antisymmetrized molecular dynamics for hypernuclei. We discuss two bands of $^{25}_{\ \Lambda}$Mg corresponding to the $K^\pi=0^+$ and $2^+$ bands of $^{24}$Mg. It is found that the energy of the $K^\pi = 2^+ \otimes \Lambda_s$ band is shifted up by about 200 keV compared to $^{24}$Mg. This is because the $\Lambda$ hyperon in $s$ orbit reduces the quadrupole deformation of the $K^\pi = 0^+ \otimes \Lambda_s$ band, while it does not change the deformation of the $K^\pi = 2^+ \otimes \Lambda_s$ band significantly.
\end{abstract}

\pacs{Valid PACS appear here}
\maketitle

\section{Introduction}

In these decades, our knowledge of $\Lambda$ hypernuclear spectra and $\Lambda N$ interaction has been greatly increased. Development of the hypernuclear gamma-ray spectroscopy has enabled us to obtain precise excitation energies \cite{Tamura_Li7L,Li7L_E,HT}. By analyzing these hypernuclear spectra theoretically and experimentally, most of the central part of $\Lambda N$ effective interaction has been clarified \cite{mill4,YNG,mill3,HT,RY,quark,mill2,Hyper}. As a consequence, it makes possible to investigate the structure of various $\Lambda$ hypernuclei systematically and quantitatively. 

By using such effective $\Lambda N$ interactions, many theoretical studies predicted and revealed unique hypernuclear phenomena caused by $\Lambda$ hyperon. For example, changes of deformation, super-symmetric state (or genuine hypernuclear state) and shrinkage of the inter-cluster distance were discussed in $p$-shell hypernuclei 
\cite{SuperSym,PSD,mill3,M_cls,light_hyp,itonaga,Li7L_T,SHFdef,def,probe,HyAMD_psd}, and some of them are confirmed by experiments \cite{Pile,Ajimura,Li7L_E,HT}. 
Recently, various structure changes have been predicted in $sd$-shell hypernuclei, because normal $sd$-shell nuclei have a variety of structure in the ground and low-lying states. 
For example, it was predicted that the $\Lambda$ hyperon in $s$ orbit reduces the nuclear deformation \cite{SHFdef,def,probe,HyAMD_psd,Lu_beta-gamma}. On the contrary, the $\Lambda$ hyperon in $p$ orbit enhances it \cite{HyAMD_psd}. 
In $^{21}_{\ \Lambda}$Ne, it was predicted that the $\Lambda$ hyperon generates various $\alpha$ + $^{16}$O + $\Lambda$ cluster states \cite{Ne21L}. By adding a $\Lambda$ particle to the $K^\pi = 2^-$ band, the mean-field like states are also generated \cite{HyAMD_Ne21L}. The difference between these structure leads to the difference in the $\Lambda$ binding energies and the reduction of the $B(E2)$ \cite{HyAMD_Ne21L}. 

Up to now, many studies have been focused on the structure change of the hypernuclei with axial deformation and/or axial symmetric cluster structure such as $^{9}_{\Lambda}$Be and $^{21}_{\ \Lambda}$Ne. Although many nuclei are considered to have axially symmetric deformation, the degree-of-freedom of triaxial deformation plays an important role in nuclei with shape coexistence, and in nuclei soft against $\gamma$ deformation \cite{Kurath_tri,Hayashi_tri,Bonche_triaxial,Redon_triaxial,Girod_Kr,Li_gsoft,Sheikh_gsoft}. 

In triaxial deformed nuclei, the response to the addition of $\Lambda$ particle will be different from those of axial symmetric nuclei \cite{Myaing_triaxial,Lu_beta-gamma,Yao_Mg25L}. 
$^{24}$Mg is one of the candidates of triaxial deformed nuclei, because of the presence of the $K^\pi = 2^+$ side band built on the $2^+_2$ state at 4.3 MeV \cite{Batchelor_Mg24,Cohen_Mg24,Girod,Bender_Mg24,Rodriguez-Egido,Mg24_AMD}. 
Based on the Skyrme-Hartree-Fock + BCS study, it was predicted that the addition of a $\Lambda$ particle makes $^{25}_{\ \Lambda}$Mg slightly soft against $\gamma$ deformation \cite{Myaing_triaxial}. And, a $\Lambda$ particle slightly stretches the ground band and reduces the $B(E2; 2^+_1 \to 0^+ )$ \cite{Yao_Mg25L}. However, in these studies, the quadrupole collective motion of $^{25}_{\ \Lambda}$Mg is treated in a approximated way by mapping the energy surface to the collective Hamiltonian. For quantitative discussions, more sophisticated treatment of deformation is desirable. Furthermore, the properties of the $K^\pi = 2^+$ band are essential for the discussion of $\gamma$ deformation, since it is quite sensitive to the $\gamma$ deformation.

The aim of the present study is to reveal how $\Lambda$ hyperon affects triaxial deformation and the observables such as excitation energy and $B(E2)$. To investigate them quantitatively, three dimensional angular momentum projection and generator coordinate method (GCM) are known to be very powerful and indispensable. Therefore we have applied the HyperAMD with the GCM (HyperAMD + GCM) to $^{25}_{\ \Lambda}$Mg. The HyperAMD \cite{HyAMD_Ne21L} is an extended version of the AMD for hypernuclei and describes hypernuclei without any assumption on the symmetry of nuclei. Combined with the GCM method, it is possible to investigate and predict the low-lying energy spectra and $B(E2)$, quantitatively. 

In this study, we focus on two positive-parity bands of $^{25}_{\ \Lambda}$Mg with a $\Lambda$ hyperon in $s$ orbit corresponding to the $K^\pi=0^+$ (ground) and $K^\pi=2^+$ bands of $^{24}$Mg, and we call the former $K^\pi=0^+ \otimes \Lambda_s$ and the latter $K^\pi=2^+ \otimes \Lambda_s$. It is found that the excitation energy of the $K^\pi=2^+ \otimes \Lambda_s$ band is shifted up by about 200 keV systematically. This is due to the difference in the binding energy of the $\Lambda$ hyperon, $B_\Lambda$, between the $K^\pi=0^+ \otimes \Lambda_s$ and $K^\pi=2^+ \otimes \Lambda_s$ bands. This difference of $B_\Lambda$ originates in the reduction of the nuclear quadrupole deformation in the $K^\pi=0^+ \otimes \Lambda_s$ band, while the $\Lambda$ hyperon does not change that in the $K^\pi=2^+ \otimes \Lambda_s$ band significantly. However, the level spacing within each band is not changed by the $\Lambda$ hyperon.

This paper organized as follows. In the next section, we explain the theoretical framework of HyperAMD + GCM.  In the Sec. III, the low-lying states with positive parity of $^{25}_{\ \Lambda}$M and their properties are discussed. The differences between the $K^\pi=0^+ \otimes \Lambda_s$ and $K^\pi=2^+ \otimes \Lambda_s$ bands are the focus. The final section summaries this work.

\section{Framework}

In this study, we applied the HyperAMD \cite{HyAMD_Ne21L} that is an extended version of AMD for hypernucleus to $^{25}_{\ \Lambda}$Mg. To analyze low-lying spectra, the generator coordinate method (GCM) was also employed.

\subsection{Hamiltonian and variational wave function}

The Hamiltonian used in this study is given as,
\begin{eqnarray}
\hat{H} &=& \hat{H}_{N} + \hat{H}_{\Lambda} - \hat{T}_g, \\
\label{H_N}
\hat{H}_{N} &=& \hat{T}_{N} + \hat{V}_{NN} + \hat{V}_{Coul},\\
\label{H_L}
\hat{H}_{\Lambda} &=& \hat{T}_{\Lambda} + \hat{V}_{\Lambda N}.
\end{eqnarray}
Here, $\hat{T}_{N}$, $\hat{T}_{\Lambda}$ and $\hat{T}_g$ are the kinetic energies of nucleons, a $\Lambda$ hyperon and the center-of-mass motion, respectively.
We have used the Gogny D1S interaction \cite{Gogny} as an effective nucleon-nucleon interaction $\hat{V}_{NN}$. The Coulomb interaction $\hat{V}_{Coul}$ is approximated by the sum of seven Gaussians.
As an effective $\Lambda N$ interaction $\hat{V}_{\Lambda N}$, we have used the central part of the YNG-NF interaction \cite{YNG}. The YNG-NF interaction depends on the nuclear Fermi momentum $k_F$ through the density dependence of the G-matrix in the nuclear medium. In this work, we apply $k_F = 1.18 $fm$^{-1}$. This gives the binding energy of $\Lambda$ particle in $^{25}_{\ \Lambda}$Mg, $B_\Lambda = 15.98$ MeV, which is consistent with the systematics of $B_\Lambda$ as a function of mass number $A$ derived from observed data \cite{Lanskoy_Yamamoto}.  

The single $\Lambda$ hypernucleus composed of $A$ nucleons and a $\Lambda$ hyperon is described by the wave function that is an eigenstate of the parity, 
\begin{eqnarray}
  \Psi^{\pi} = \hat{P}^{\pi}\Psi_{\rm int},
\end{eqnarray}
where $\hat{P}^{\pi}$ is the parity projector and the intrinsic wave function $\Psi_{\rm int}$ is represented by the direct product of the $\Lambda $ single-particle wave function $\varphi_\Lambda$ and the wave function of $^{24}$Mg, $\Psi_N$, 
\begin{eqnarray}
\Psi_{\rm int} &=& \varphi_\Lambda \otimes \Psi_N.
\label{int_wf}
\end{eqnarray}
The nuclear part is described by a Slater determinant of nucleon single-particle
wave packets,
\begin{eqnarray}
\Psi_N &=& \frac{1}{\sqrt{A!}}\det \left\{ \psi_{i} \left( r_j \right) \right\},\\
\psi_{i} \left( r_j \right) &=& \phi_{i} \left( r_j \right) \cdot \chi_{i}\cdot \eta_{i},\\
\phi_{i} \left( r   \right) &=& \prod_{\sigma=x,y,z} \biggl(\frac{2\nu_\sigma}{\pi}\biggr)^{1/4}
 \exp \biggl\{-\nu_\sigma \bigl(r - Z_{i} \bigr)_\sigma^2 \biggr\},\\ 
\chi_{i} &=& \alpha_i \chi_\uparrow + \beta_i \chi_\downarrow,\\
\eta_{i} &=& {\rm proton}\ {\rm or}\ {\rm neutron},
\end{eqnarray}
where $\psi_{i}$ is {\it i}th nucleon single-particle wave packet consisting of spatial
$\phi_{i}$, spin $\chi_{i}$ and isospin $\eta _{i}$ parts. 
The centroids of Gaussian $\bm{Z}_i$, width parameters $\nu_\sigma$ and spin directions $\alpha_i$ and $\beta_i$ are the variational parameters of the nuclear part. 

The $\Lambda$ single-particle wave function is represented by the superposition of  Gaussian wave packets,
\begin{eqnarray}
\varphi_\Lambda \left( r \right) = \sum_{m=1}^M c_m \varphi_m \left( r \right),\quad 
\varphi_m \left( r \right) = \phi_m \left( r \right) \cdot \chi_m,\\
\phi_m \left( r \right) = \prod_{\sigma=x,y,z} \biggl(\frac{2\nu_\sigma \mu}{\pi}\biggr)^{1/4}
 \exp \biggl\{-\nu_\sigma \mu \bigl(r - z_m \bigr)_\sigma^2 \biggr\},
\end{eqnarray}
\begin{eqnarray}
\chi_m &=& a_m \chi_\uparrow + b_m \chi_\downarrow, \\
\mu &=& \frac{m_\Lambda}{m_N},
\end{eqnarray}
where $m_\Lambda$ and $m_N$ represent the masses of the $\Lambda$ particle and the nucleon, respectively.
Again, the centroid of Gaussian $\bm{z}_m$, spin directions $a_m$ and $b_m$, and coefficients $c_m$ are the variational parameters of the hyperon part.
Since we have superposed Gaussian wave packets, it is rather tedious to remove the spurious center-of-mass kinetic energy exactly. Therefore we approximately removed it in the same way as our previous work \cite{HyAMD_Ne21L}.

\subsection{Variation on $\beta$-$\gamma$ plane}

The variational calculation has been performed in two steps. The first is the variational calculation  for $^{24}$Mg under the constraints on nuclear matter quadrupole deformation parameters $\beta$ and $\gamma$. The $\beta$-$\gamma$ constraint is applied in the same way to the references \cite{Mg24_AMD,Suhara_bg}. 
The variation with the $\beta$-$\gamma$ constraint was achieved by addition of the parabolic potentials,
\begin{eqnarray}
 V_c &=& v_\beta \left( \langle \beta \rangle - \beta_i \right)^2 + v_\gamma \left( \langle \gamma \rangle - \gamma_i \right)^2,\\
 \tilde{E} &=& \frac{\langle \Psi^\pi_N | \hat{H} | \Psi^\pi_N \rangle}{\langle \Psi^\pi_N | \Psi^\pi_N \rangle} + V_c,
 \label{pot}
\end{eqnarray}
to the total energy of the core nucleus $^{24}$Mg. The variational wave function of the nucleon part was determined to minimize $\tilde{E}$ for given values of $\beta_i$ and $\gamma_i$ by using the frictional cooling method. The resulting nuclear wave function $\Psi^\pi_N (\beta_i,\gamma_i )$ has minimum energy for given set of $(\beta_i,\gamma_i )$ and we shall call it core state. In this study, the core states are calculated for discrete sets of $(\beta_i, \gamma_i )$ in $\beta$-$\gamma$ plane. 

The second step is the variation for $^{25}_{\ \Lambda}$Mg. Combined with the core-state wave function $\Psi^\pi_N (\beta_i, \gamma_i )$ as $ \Psi^\pi (\beta_i, \gamma_i) = \varphi_\Lambda \otimes \Psi^\pi_N \left( \beta_i, \gamma_i \right) $, we have performed variational calculation of the $\Lambda$ single-particle wave function and determined the variational parameters $\bm{z}_m$, $a_m$, $b_m$ and $c_m$ for each grid point on $\beta$-$\gamma$ plane. We shall call the resulting state $\Psi^\pi (\beta_i, \gamma_i)$ hypernuclear state.

\subsection{Angular momentum projection and GCM}

After the variational calculation, we project out an eigenstate of the total angular momentum from the hypernuclear states, 
\begin{eqnarray}
 \label{AngProj}
 \Psi^{J\pi}_{MK}(\beta_i,\gamma_i) &=& \hat{P}^{J}_{MK} \Psi^{\pi} (\beta_i,\gamma_i)/ \sqrt{ N^{J\pi}_{K}(\beta_i,\gamma_i) } ,
\end{eqnarray}
\begin{eqnarray}
 \hat{P}^{J}_{MK} &=& \frac{2J+1}{8\pi^2} \int d\Omega D^{J*}_{MK}(\Omega) \hat{R}(\Omega),\\
 N^{J\pi}_{K} &=& \langle \hat{P}^{J}_{MK} \Psi^{\pi} (\beta_i,\gamma_i) | \hat{P}^{J}_{MK} \Psi^{\pi} (\beta_i,\gamma_i) \rangle.
\end{eqnarray}
Here $\hat{P}^{J}_{MK}$ is the total angular momentum projector. The integrals are performed numerically over three Eular angles $\Omega$.

We calculate the mixing between the different $K$ states that have the same intrinsic deformation $( \beta_i, \gamma_i )$,
\begin{eqnarray}
\label{Kmixing}
\Psi^{J\pi}_{n}\left( \beta_i, \gamma_i \right) = \sum_{K=-J}^{J} f_{K n i} \Psi^{J\pi}_{MK} \left( \beta_i, \gamma_i \right),
\end{eqnarray}
and call it $K$-mixed state.
The coefficients $f_{Kni}$ are determined through the double diagonalization of $H_{KK'}$ and $N_{KK'}$ defined as,
\begin{eqnarray}
H_{KK'} = \langle \Psi^{J \pi}_{MK} (\beta_i, \gamma_i) | \hat{H} | \Psi^{J \pi}_{MK'} (\beta_i, \gamma_i) \rangle,\\
N_{KK'} = \langle \Psi^{J \pi}_{MK} (\beta_i, \gamma_i) | \Psi^{J \pi}_{MK'} (\beta_i, \gamma_i) \rangle.
\end{eqnarray}

Finally, we superpose all of the $K$-mixed states with different deformation ( $\beta_i$ , $\gamma_i $ ) (GCM).
Then the final wave function of the system becomes as follows: 
\begin{eqnarray}
\Psi^{J\pi}_{\alpha} &=& \sum_{i,n} g_{n i \alpha} \Psi^{J\pi}_{n}\left( \beta_i, \gamma_i \right),
\label{GCM_wf}
\end{eqnarray}
where quantum numbers other than total angular momentum and parity are represented by $\alpha$. 
The coefficients $g_{ni\alpha}$ are determined by the solving the Hill-Wheeler equation: 
\begin{eqnarray}
\sum_{n',j} H_{n i n' j} g_{n' j \alpha} &=& E_\alpha \sum_{n',j} N_{n i n' j} g_{n' j \alpha},
\end{eqnarray}
\begin{eqnarray}
H_{nin'j} &=& \langle \Psi^{J \pi}_{n} (\beta_i, \gamma_i) |\hat{H} | \Psi^{J \pi}_{n'} (\beta_j, \gamma_j) \rangle,\\
N_{nin'j} &=& \langle \Psi^{J \pi}_{n} (\beta_i, \gamma_i) | \Psi^{J \pi}_{n'} (\beta_j, \gamma_j) \rangle.
\end{eqnarray}
The physical quantities discussed in the next section are basically calculated from the GCM wave function.

\subsection{Analysis of wave function}

To analyze and discuss the GCM wave function on $\beta$-$\gamma$ plane, it is convenient to introduce the overlap between the GCM wave function $\Psi^{J\pi}_n$ and $K$-mixed states, 
\begin{eqnarray}
O^{J\pi}_{n \alpha} (\beta_i, \gamma_i ) = | \langle \Psi^{J \pi}_{n} (\beta_i, \gamma_i ) | \Psi^{J \pi}_\alpha \rangle |^2.
\label{Overlap}
\end{eqnarray} 
The behavior of $O^{J\pi}_{n \alpha} (\beta_i, \gamma_i )$ in $\beta$-$\gamma$ plane is discussed in the next section.

We also calculate the expectation values of the operators $\hat{H}_\Lambda$ and $\hat{H}_N$,
\begin{eqnarray}
\label{B_Lmd}
B_\Lambda &=& -\langle \Psi^{J \pi}_{\alpha} | \hat{H}_\Lambda | \Psi^{J \pi}_{\alpha} \rangle,\\
\label{E_N}
E_N &=& \langle \Psi^{J \pi}_{\alpha} | \hat{H}_N | \Psi^{J \pi}_{\alpha} \rangle,
\end{eqnarray}
to see the contribution from the hyperon part and nuclear part to energy shifts.
For further analysis, by using the $K$-mixed state Eq. (\ref{Kmixing}) with given deformation parameters $\beta_i$ and $\gamma_i$, we calculate the expectation values of the operators $\hat{H}$ and $\hat{H}_\Lambda$,
\begin{eqnarray}
\label{KmixedE}
\epsilon \left( \beta_i , \gamma_i \right) &=& \langle \Psi^{J\pi}_{n}\left( \beta_i, \gamma_i \right) | \hat{H} | \Psi^{J\pi}_{n}\left( \beta_i, \gamma_i \right) \rangle,\\
\label{b_bg}
b_\Lambda \left( \beta_i , \gamma_i \right) &=& -\langle \Psi^{J\pi}_{n}\left( \beta_i, \gamma_i \right) | \hat{H}_\Lambda | \Psi^{J\pi}_{n}\left( \beta_i, \gamma_i \right) \rangle.
\end{eqnarray}

\section{Results and discussions}

\begin{figure*}
  \begin{center}
    \includegraphics[keepaspectratio=true,width=172mm]{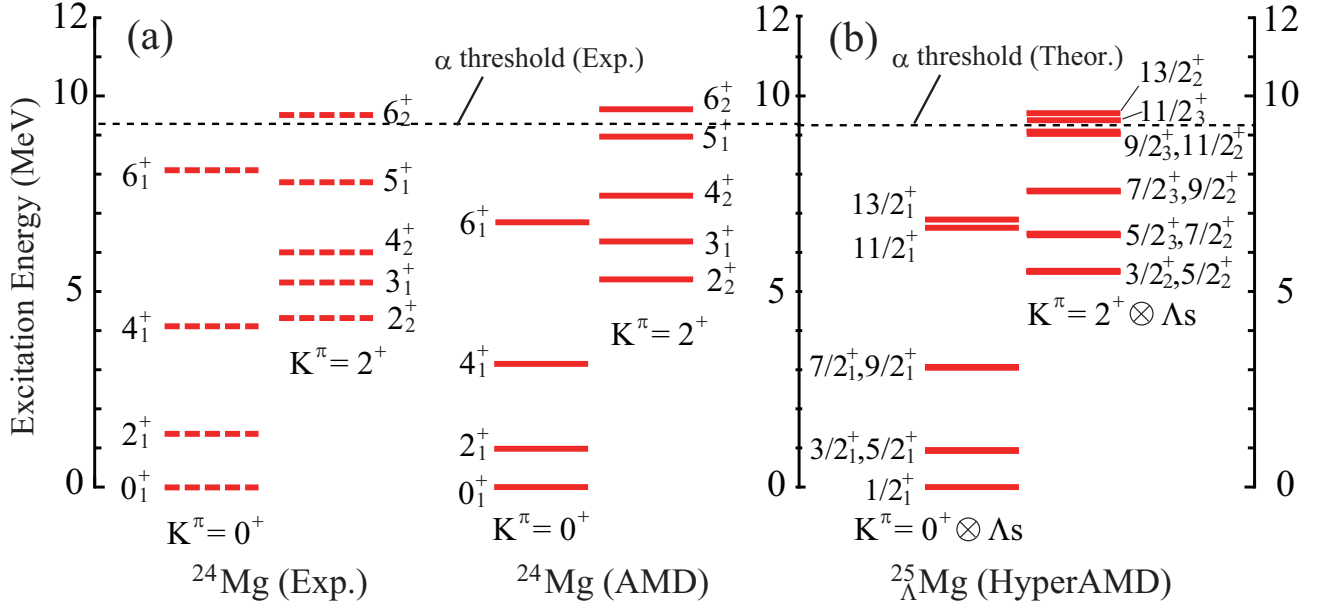}
  \end{center}
  \caption{(Color online) (a) Calculated and observed low-lying energy spectra of $^{24}$Mg. (b) Corresponding spectra of $^{25}_{\ \Lambda}$Mg with $\Lambda$ hyperon in $s$ orbit. }
  \label{fig:levels.eps}
\end{figure*}

\begin{figure}
  \begin{center}
    \includegraphics[keepaspectratio=true,width=86mm]{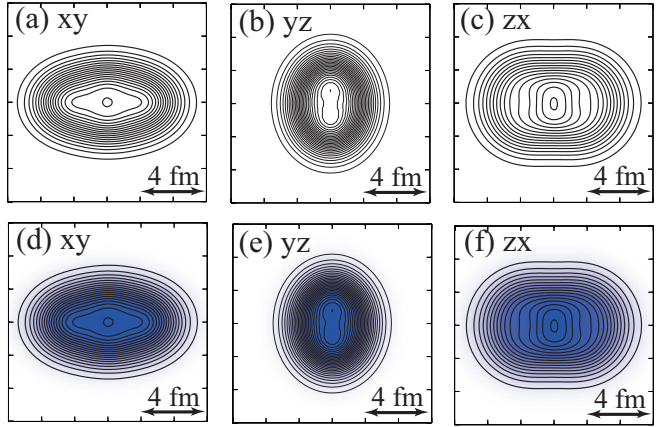}
  \end{center}
  \caption{(Color online) Density distributions of the intrinsic wave functions that contribute largely to the $2^+_1$ and $2^+_2$ states of $^{24}$Mg (upper panels) and to the $3/2^+_1$ and $3/2^+_2$ states of $^{25}_{\ \Lambda}$Mg (bottom panels) plotted from different directions, respectively. Solid lines show the nuclear density distributions, while the color plots represent the distributions of the $\Lambda$ hyperon.}
  \label{fig:dens.eps}
\end{figure}

The calculated and observed energy spectra of $^{24}$Mg are presented in Fig. \ref{fig:levels.eps}(a). The AMD with GCM framework describes successfully the ground $(K^\pi = 0^+ )$ and the $K^\pi = 2^+$ bands. The dotted-line at 9.32 MeV represents the experimental $\alpha + ^{20}$Ne threshold energy corresponding to the lowest decay channel. 
In $^{24}$Mg, a triaxial deformation of low-lying states has been discussed by many authors \cite{Batchelor_Mg24,Cohen_Mg24,Girod,Bender_Mg24,Rodriguez-Egido,Mg24_AMD}. The excitation energy of the $K^\pi=2^+$ band is quite sensitive to the degree-of-freedom of triaxial deformation. If triaxial deformation is not included, the theoretical calculations overestimate the excitation energy of the $2^+_2$ state by about 4 MeV \cite{Bender_Mg24,Mg24_AMD}. 
Fig. \ref{fig:dens.eps}(a)-(c) display the density distributions of the intrinsic wave functions which contribute largely to the $0^+_1$ and $2^+_2$ states of $^{24}$Mg. It confirms triaxial deformation of $^{24}$Mg.

\subsection{Energy spectra of $^{25}_{\ \Lambda}$M\lowercase{g}}

Fig. \ref{fig:levels.eps}(b) shows the low-lying spectrum of $^{25}_{\ \Lambda}$Mg. We focus on two bands corresponding to the $K^\pi=0^+$ and $K^\pi=2^+$ bands of $^{24}$Mg generated by adding a $\Lambda$ hyperon in $s$ orbit. We call the former $K^\pi=0^+ \otimes \Lambda_s$ band, and the latter $K^\pi=2^+ \otimes \Lambda_s$ band. Coupling of a $\Lambda$ hyperon in $s$ orbit to the non-zero $J$ states generates the doublets with $J - 1/2$ and $J + 1/2$. However, most of them are degenerated in Fig. \ref{fig:levels.eps}. 
The dotted line in Fig. \ref{fig:levels.eps}(b) represents the $\alpha + ^{21}_{\ \Lambda}$Ne threshold energy calculated with HyperAMD.
It shows that both the $K^\pi=0^+ \otimes \Lambda_s$ and $K^\pi=2^+ \otimes \Lambda_s$ bands will bound.

Fig. \ref{fig:dens.eps}(d)-(f) display the intrinsic density distributions for the $3/2^+_2$ state of $^{25}_{\ \Lambda}$Mg belonging to the $K^\pi=2^+ \otimes \Lambda_s$ band. It shows that the nuclear density distribution of the $3/2^+_2$ state in $^{25}_{\ \Lambda}$Mg has triaxial deformation.  Therefore, the contribution from the intrinsic state with triaxial deformation is important in $^{25}_{\ \Lambda}$Mg as in the case of $^{24}$Mg. 
The distribution of the $\Lambda$ hyperon is also triaxialy deformed. Therefore the $\Lambda$ hyperon orbit is not pure $s$ orbit but non-zero angular momentum components are also mixed. In this paper, we call it $s$ orbit approximately. 

Although the drastic changes cannot be seen in Fig. \ref{fig:levels.eps}, the $\Lambda$ hyperon changes the excitation energies quantitatively.
In Tab. \ref{Tab:spectra}, the excitation energies of each state in $^{24}$Mg and $^{25}_{\ \Lambda}$Mg are listed. It shows that the excitation energy of the $K^\pi=2^+ \otimes \Lambda_s$ band is shifted up by about $\Delta E_x = 200$ keV systematically. 
On the other hand, the level spacing within the $K^\pi=0^+ \otimes \Lambda_s$ and $K^\pi=2^+ \otimes \Lambda_s$ bands hardly change by a $\Lambda$ hyperon.
This is different from the prediction by reference \cite{Yao_Mg25L}, where the authors predicted that the $\Lambda$ hyperon slightly stretches the spectra of the ground band, due to the reduction of the $\beta$ deformation. 

\begin{table}
  \caption{Excitation energies (MeV) of each state in the $K^\pi=0^+$ and $K^\pi=2^+$ bands of $^{24}$Mg and the corresponding states of $^{25}_{\ \Lambda}$Mg. Changes of excited energies $\Delta E_x$ for each state obtained with AMD are also presented in unit of MeV. For comparison, observed energies are also listed \cite{A21_A44}.}
  \label{Tab:spectra}
  \begin{ruledtabular}
  \begin{tabular}{cccccc}
  \multicolumn{3}{c}{$^{24}$Mg} & 
  \multicolumn{2}{c}{$^{25}_{\ \Lambda}$Mg} & \\
  \cline{1-3}\cline{4-5}
  $K^\pi=0^+$  & $E_x$(Cal.) & $E_x$(Exp.) & $0^+ \otimes \Lambda_s$ & $E_x$(Cal.) & $\Delta E_x$ \\
  \hline
  $0^+_1$ & 0.00 & 0.00 & $1/2^+_1$ & 0.00 & $\pm$0.00 \\
  $2^+_1$ & 0.98 & 1.37 & $3/2^+_1$ & 0.94 & $-$0.04 \\
          &      &      & $5/2^+_1$ & 0.93 & $-$0.05 \\
  $4^+_1$ & 3.15 & 4.12 & $7/2^+_1$ & 3.06 & $-$0.09 \\
          &      &      & $9/2^+_1$ & 3.06 & $-$0.09 \\
  $6^+_1$ & 6.77 & 8.11 & $11/2^+_1$ & 6.63 & $-$0.14 \\
          &      &      & $13/2^+_1$ & 6.84 & $+$0.07 \\
  \\
  \multicolumn{3}{c}{$^{24}$Mg} & 
  \multicolumn{2}{c}{$^{25}_{\ \Lambda}$Mg} & \\
  \cline{1-3}\cline{4-5}
  $K^\pi=2^+$ & $E_x$(Cal.) & $E_x$(Exp) & $2^+ \otimes \Lambda_s$ & $E_x$(Cal.) & $\Delta E_x$ \\
  \hline
  $2^+_2$ & 5.31 & 4.33 & $3/2^+_2$ & 5.53 & $+$0.22 \\
          &      &      & $5/2^+_2$ & 5.50 & $+$0.19 \\
  $3^+_1$ & 6.28 & 5.24 & $5/2^+_3$ & 6.44 & $+$0.16 \\
          &      &      & $7/2^+_2$ & 6.48 & $+$0.20 \\
  $4^+_2$ & 7.45 & 6.01 & $7/2^+_3$ & 7.56 & $+$0.11 \\
          &      &      & $9/2^+_2$ & 7.58 & $+$0.13 \\
  $5^+_1$ & 8.96 & 7.81 & $9/2^+_3$ & 9.03 & $+$0.07 \\
          &      &      & $11/2^+_2$ & 9.09 & $+$0.13 \\
  $6^+_2$ & 9.66 & 9.53 & $11/2^+_3$ & 9.38 & $-$0.22 \\
          &      &      & $13/2^+_2$ & 9.56 & $-$0.10 \\
  \end{tabular}
  \end{ruledtabular}
\end{table}

\subsection{Difference in $B_\Lambda$ between the \bm{$K^\pi = 0^+ \otimes \Lambda_s$} and \bm{$K^\pi = 2^+ \otimes \Lambda_s$} bands}

\begin{figure}
  \begin{center}
    \includegraphics[keepaspectratio=true,width=86mm]{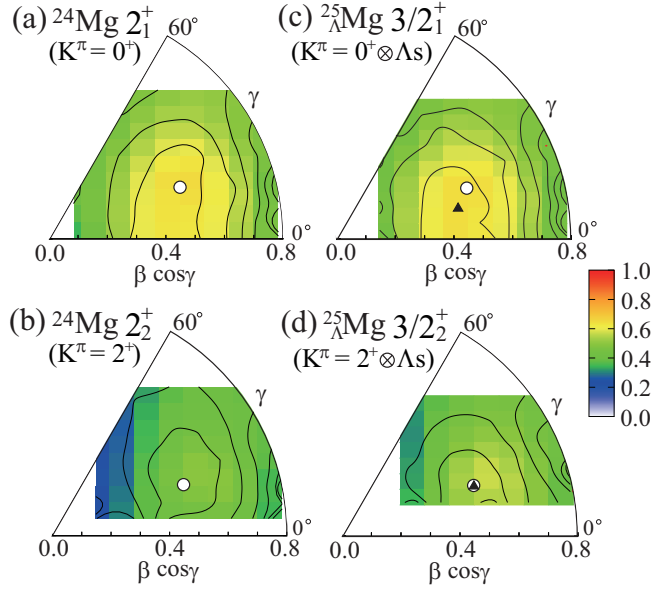}
  \end{center}
  \caption{(Color online) Contour line represents the increase of $\epsilon (\beta, \gamma)$ for every 1 MeV, and color plots show the distribution of the GCM overlap. The definition of $\epsilon (\beta, \gamma)$ and the GCM overlap are given by Eq. (\ref{KmixedE}) and Eq. (\ref{Overlap}), respectively. (a) and (b) correspond to the $2^+_1$ states of the $K^\pi=0^+$ band and the $2^+_2$ state of the $K^\pi = 2^+ $ band of $^{24}$Mg, respectively. Open circles represent the peak positions of the GCM overlap for the $2^+_1$ and $2^+_2$ states, respectively.  (c) and (d) correspond to the $3/2^+_1$ and $3/2^+_2$ states of $^{25}_{\ \Lambda}$Mg with $\Lambda$ hyperon in $s$ orbit, respectively. Filled triangles show the peak positions of the GCM overlap for each state of $^{25}_{\ \Lambda}$Mg, while open circles show those of the GCM overlap for the corresponding states $2^+_1$ and $2^+_2$, respectively.}
  \label{fig:Energy-GCMamp.eps}
\end{figure}

\begin{table}
  \caption{Total energy $E$, changes of $E_N$, $\Delta E_N$, and $B_\Lambda$ are listed. Definition of $\Delta E_N$ is given in text.}
  \label{Tab:components}
  \begin{ruledtabular}
  \begin{tabular}{cccc}
  \multicolumn{4}{c}{$K^\pi = 0^+ \otimes \Lambda_s$ band} \\
  States & $E$ & $\Delta E_N$ & $B_\Lambda$  \\
  \hline
  $1/2^+_1$ & -213.83  & +0.08  & 15.98  \\
  $3/2^+_1$ & -212.89  & +0.04  & 15.99  \\
  $5/2^+_1$ & -212.89  & +0.04  & 15.99  \\
  $7/2^+_1$ & -210.77  & +0.02  & 16.03  \\
  $9/2^+_1$ & -210.76  & +0.04  & 16.04  \\
  \\
  \multicolumn{4}{c}{$K^\pi = 2^+ \otimes \Lambda_s$  band} \\
  States & $E$ & $\Delta E_N$ & $B_\Lambda$ \\
  \hline
  $3/2^+_2$ & -208.30  & +0.10  & 15.80  \\  
  $5/2^+_2$ & -208.33  & +0.07  & 15.79  \\
  $5/2^+_3$ & -207.39  & +0.03  & 15.79  \\
  $7/2^+_2$ & -207.35  & +0.07  & 15.79  \\
  $7/2^+_3$ & -206.27  & +0.01  & 15.82  \\
  $9/2^+_2$ & -206.25  & +0.04  & 15.82  \\
  \end{tabular}
  \end{ruledtabular}
\end{table}

The increase of the excitation energy of the $K^\pi=2^+ \otimes \Lambda_s$ band is due to the difference in the $\Lambda$ binding energy $B_\Lambda$ between the $K^\pi=0^+ \otimes \Lambda_s$ and $K^\pi=2^+ \otimes \Lambda_s$ bands.
Tab. \ref{Tab:components} summarizes the $B_\Lambda$ defined by Eq. (\ref{B_Lmd}) for each hypernuclear state in the $K^\pi=0^+ \otimes \Lambda_s$ and $K^\pi=2^+ \otimes \Lambda_s$ bands. 
For comparison, we calculated the energy changes of the nuclear part, $\Delta E_N$, as, 
\begin{eqnarray}
\Delta E_N &=& E_N\left(^{25}_{\ \Lambda}\mathrm{Mg}(J^\pi) \right) - E\left(^{24}\mathrm{Mg}(j^\pi) \right),
\label{deltaEN}
\end{eqnarray}
where $ E_N$ is defined by Eq. (\ref{E_N}).
The total energy $E$ and $\Delta E_N$ obtained after the GCM calculation are also listed in Tab. \ref{Tab:components}. It shows that the $B_\Lambda$ for the $K^\pi=0^+ \otimes \Lambda_s$ band is larger than that for the $K^\pi=2^+ \otimes \Lambda_s$ band by 200 keV systematically. On the other hand, the effects of the $\Lambda$ particle to the nuclear part are small and comparable for the $K^\pi=0^+$ and $K^\pi=2^+$ bands. 
Combining the Eqs. (\ref{B_Lmd}), (\ref{E_N}) and (\ref{deltaEN}), one finds the relation,
\begin{eqnarray}
\Delta E_x &=& E_x\left(^{25}_{\ \Lambda}\mathrm{Mg}(J^\pi) \right) - E_x\left(^{24}\mathrm{Mg}(j^\pi) \right) \\
&=& \Delta E_N - B_\Lambda + B_\Lambda(GS), \\
B_\Lambda (GS) &=& E\left( ^{24}\mathrm{Mg}(GS) \right) - E\left(^{25}_{\ \Lambda}\mathrm{Mg}(GS) \right),
\end{eqnarray}
where $\Delta E_x $ is change of excitation energy between $^{24}$Mg and $^{25}_{\ \Lambda}$Mg for each state and we have obtained the $B_\Lambda (GS) = 15.91$ MeV. 
Therefore the difference of the $\Delta E_x$ between the $K^\pi=0^+ \otimes \Lambda_s$ and $K^\pi=2^+ \otimes \Lambda_s$ bands comes from the difference of $B_\Lambda$ between these bands.

The difference of $B_\Lambda$ between the $K^\pi=0^+ \otimes \Lambda_s$ and $K^\pi=2^+ \otimes \Lambda_s$ bands comes from the difference of the deformation change by $\Lambda$ hyperon between these bands. The deformation change is clearly seen from the distribution of the GCM overlap defined by Eq. (\ref{Overlap}). In Fig. \ref{fig:Energy-GCMamp.eps}, the color plots display the distribution of the GCM overlap for the $2^+_1$ and $2^+_2$ states of $^{24}$Mg and for the $3/2^+_1$ and $3/2^+_2$ states of $^{25}_{\ \Lambda}$Mg. And the contour shows the increase of $\epsilon ( \beta, \gamma) $, defined by Eq. (\ref{KmixedE}). To obtain the $\epsilon ( \beta, \gamma)$, we perform the diagonalization of $K$ and obtain two eigenvalues for the $J=2$ state of $^{24}$Mg and the $J=3/2$ state of $^{25}_{\ \Lambda}$Mg, respectively. It is found that the lowest energy eigenstate corresponds to the $2^+_1$ ($3/2^+_1$) state of $^{24}$Mg ($^{25}_{\ \Lambda}$Mg), and the second lowest eigenstate corresponds to the $2^+_2$ ($3/2^+_2$) state of $^{24}$Mg ($^{25}_{\ \Lambda}$Mg) at each $( \beta_i, \gamma_i )$. 

In Fig. \ref{fig:Energy-GCMamp.eps}(a), the peak of the GCM overlap for the $2^+_1$ state of $^{24}$Mg locates at $(\beta, \gamma) = ( 0.48, 21^\circ )$. Fig. \ref{fig:Energy-GCMamp.eps}(b) shows that the peak of the GCM overlap also locates at $(\beta, \gamma) = ( 0.48, 21^\circ )$ for the $2^+_2$ state. In the case of the $3/2^+_1$ state of $^{25}_{\ \Lambda}$Mg, the peak position is shifted toward smaller $\beta$ and $\gamma$ compared to $^{24}$Mg, and located at $(\beta, \gamma) = ( 0.43, 15^\circ )$. On the contrary, in the case of the $3/2^+_2$ state, the peak position is unchanged. Therefore the $\Lambda$ hyperon in $s$ orbit reduces the quadrupole deformation of the $3/2^+_1$ state, while it does not change deformation of the $3/2^+_2$ state.

Here we compare the $b_\Lambda ( \beta, \gamma )$ of the $3/2^+_1$ state with that of the $3/2^+_2$ state in $^{25}_{\ \Lambda}$Mg. According to Eq. (\ref{b_bg}), we obtained the $b_\Lambda ( \beta=0.43, \gamma=15^\circ ) = 16.09$ MeV for the $3/2^+_1$ state, while $b_\Lambda ( \beta=0.48, \gamma=21^\circ ) = 15.80$ MeV for the $3/2^+_2$ state. Namely, the $b_\Lambda (\beta, \gamma)$ of the $3/2^+_1$ state is larger than that of the $3/2^+_2$ state. This is because the nuclear quadrupole deformation of the $3/2^+_1$ state is smaller than that of the $3/2^+_2$ state. Therefore the $\Lambda$ hyperon in $s$ orbit gains the larger $\Lambda$ binding energy for the $3/2^+_1$ state. 
This trend is common to the other excited states. Deformation of the member states of the $K^\pi = 0^+ \otimes \Lambda_s$ band is always reduced and those of the $K^\pi = 2^+ \otimes \Lambda_s$ band is rarely changed (Tab. \ref{Tab:bet-gam}). It leads to the systematic reduction of $B_\Lambda$ in the $K^\pi = 2^+ \otimes \Lambda_s$ band. As a results, the $K^\pi = 2^+ \otimes \Lambda_s$ band is shifted up by about 200 keV.

\begin{table}
  \caption{Deformation parameters $\beta$ and $\gamma$ corresponding to the peak of the GCM overlap are summarized. $b_\Lambda (\beta, \gamma)$ at the peak position of the GCM overlap for each state is also listed in unit of MeV. Definition of $ b_\Lambda (\beta, \gamma)$ is given by Eq. (\ref{b_bg}).}
  \label{Tab:bet-gam}
  \begin{ruledtabular}
  \begin{tabular}{ccccccc}
  \multicolumn{3}{c}{$^{24}$Mg} & 
  \multicolumn{4}{c}{$^{25}_{\ \Lambda}$Mg} \\
  \cline{1-3}\cline{4-7}
  $K^\pi=0^+$  & $\beta$ & $\gamma$ & $0^+ \otimes \Lambda_s$ & $\beta$ & $\gamma$ & $b_\Lambda$ \\
  \hline
  $0^+_1$ & 0.48  & 21$^\circ$  & $1/2^+_1$ & 0.48  & 21$^\circ$  & 15.90 \\
  $2^+_1$ & 0.48  & 21$^\circ$  & $3/2^+_1$ & 0.43  & 15$^\circ$  & 16.09 \\ 
          &       &             & $5/2^+_1$ & 0.43  & 15$^\circ$  & 16.10 \\
  $4^+_1$ & 0.48  & 21$^\circ$  & $7/2^+_1$ & 0.39  & 25$^\circ$  & 16.14 \\ 
          &       &             & $9/2^+_1$ & 0.39  & 25$^\circ$  & 16.15 \\
  \\
    \multicolumn{3}{c}{$^{24}$Mg} & 
  \multicolumn{4}{c}{$^{25}_{\ \Lambda}$Mg} \\
  \cline{1-3}\cline{4-7}
  $K^\pi=2^+$  & $\beta$ & $\gamma$ & $2^+ \otimes \Lambda_s$ & $\beta$ & $\gamma$ & $b_\Lambda$ \\
  \hline
  $2^+_2$ & 0.48  & 21$^\circ$  & $3/2^+_2$ & 0.48  & 21$^\circ$  & 15.80 \\
          &       &             & $5/2^+_2$ & 0.48  & 21$^\circ$  & 15.80 \\
  $3^+_1$ & 0.48  & 21$^\circ$  & $5/2^+_3$ & 0.48  & 21$^\circ$  & 15.79 \\ 
          &       &             & $7/2^+_2$ & 0.48  & 21$^\circ$  & 15.79 \\
  $4^+_2$ & 0.48  & 21$^\circ$  & $7/2^+_3$ & 0.48  & 21$^\circ$  & 15.77 \\ 
          &       &             & $9/2^+_2$ & 0.48  & 21$^\circ$  & 15.78 \\
  \end{tabular}
  \end{ruledtabular}
\end{table}

\subsection{Changes of $B(E2)$ by $\Lambda$ hyperon}

We have calculated the intra- and inter- band $B(E2)$ values by using the GCM wave functions. To compare the $B(E2)$ values of $^{25}_{\ \Lambda}$Mg with those of $^{24}$Mg, they are corrected under the assumption that a $\Lambda$ hyperon occupies the $s$ orbit for each hypernuclear state in the $K^\pi=0^+ \otimes \Lambda_s$ and $K=2^+ \otimes \Lambda_s $ bands (see the Appendix of the reference \cite{HyAMD_Ne21L}). Tab. \ref{Tab:E2} summarizes both the uncorrected and corrected $B(E2)$ values corresponding to the intra-band transitions $2^+_1 \to 0^+_1$ in the $K^\pi=0^+$ band and $3^+_1 \to 2^+_2$ in the $K^\pi=2^+$ band. 

It is found that the $\Lambda$ hyperon in $s$ orbit does not change the $B(E2)$ values significantly. Indeed, Tab. \ref{Tab:E2} shows that the changes of $B(E2)$ due to the $\Lambda$ hyperon are less than 10 \% except for the $7/2^+_2 \to 3/2^+_2$ and $5/2^+_3 \to 5/2^+_2$ transitions, and are consistent with the prediction by reference \cite{Yao_Mg25L}. These changes are smaller than the changes in $^{21}_{\ \Lambda}$Ne. In the case of $^{21}_{\ \Lambda}$Ne, it is predicted that the $B(E2)$ values are reduced by about 20 \% by the calculation with the $\alpha + ^{16}$O $+ \Lambda$ cluster model \cite{Ne21L} and HyperAMD + GCM \cite{HyAMD_Ne21L}. We have calculated and investigated the change of $B(E2)$ values for the inter-band transitions in $^{25}_{\ \Lambda}$Mg. It is found that the changes of $B(E2)$ values of the intra-band transitions are also about 10\%.

\begin{table}
\caption{Intra-band $B(E2)$ values ($e^2$fm$^4$) for the $ 2^+_1 \to 0^+_1 $ ($K^\pi=0^+$ band) and $ 3^+_1 \to 2^+_2 $ ($K^\pi=2^+$ band) transitions in $^{24}$Mg and the cprresponding transitions in $^{25}_{\ \Lambda}$Mg. $cB(E2)$ represents the corrected $B(E2)$ values obtained by the same way as explained in the reference \cite{HyAMD_Ne21L}.}
\label{Tab:E2}
  \begin{ruledtabular}
  \begin{tabular}{cccccc}
  \multicolumn{2}{c}{$^{24}$Mg (AMD)} & \multicolumn{3}{c}{$^{25}_{\ \Lambda}$Mg (HyperAMD)} &  \\
  \cline{1-2} \cline{3-5}
  Trasitions & $B(E2)$ & Transitions & $B(E2)$ & $cB(E2)$ & Changes(\%) \\
  \hline
  $2^+_1 \to 0^+_1$ & 98  & $3/2^+_1 \to 1/2^+_1$ &  92 &  92 & $-$6.1 \\
                    &     & $5/2^+_1 \to 1/2^+_1$ &  92 &  92 & $-$6.1 \\
  $3^+_1 \to 2^+_2$ & 167 & $7/2^+_2 \to 3/2^+_2$ &  19 & 192 & $+$15 \\
                    &     & $7/2^+_2 \to 5/2^+_2$ & 160 & 178 & $+$6.6 \\
                    &     & $5/2^+_3 \to 3/2^+_2$ & 138 & 173 & $+$3.4 \\
                    &     & $5/2^+_3 \to 5/2^+_2$ & 39 & 195 & $+$17 \\
  \end{tabular}

  \end{ruledtabular}
\end{table}

\section{Summary} 

In this paper, we have applied the HyperAMD to $^{25}_{\ \Lambda}$Mg and investigated the effects by a $\Lambda$ hyperon to the nucleus with triaxial deformation. Focusing on the positive-parity states with a $\Lambda$ hyperon in $s$ orbit, we discussed two bands of $^{25}_{\ \Lambda}$Mg corresponding to the $K^\pi=0^+$ and $K^\pi=2^+$ bands in $^{24}$Mg. Both of them are expected to be bound. 

Although the gross feature of the excitation spectra of $^{25}_{\ \Lambda}$Mg remains similar to that of $^{24}$Mg, the $\Lambda$ hyperon in $s$ orbit changes the excitation energies quantitatively.
It is found that the excitation energy of the $K^\pi=2^+ \otimes \Lambda_s $ band in $^{25}_{\ \Lambda}$Mg is shifted up by 200 keV compared to that of the $K^\pi = 2^+$ band in $^{24}$Mg. This comes from the difference of $B_\Lambda$ between the $K^\pi=0^+ \otimes \Lambda_s$ and $K^\pi=2^+ \otimes \Lambda_s$ bands and it depends on the deformation change of these bands. Since the $\Lambda$ hyperon reduces the nuclear quadrupole deformation of the $ K^\pi=0^+ \otimes \Lambda_s $ band, the $\Lambda$ binding energy of the $K^\pi=0^+ \otimes \Lambda_s$ is larger than that of the $K^\pi=2^+ \otimes \Lambda_s$ band. 
On the other hand, the level spacing with the $K^\pi=0^+ \otimes \Lambda_s$ and $K^\pi=2^+ \otimes \Lambda_s$ bands does not change by adding a $\Lambda$ hyperon significantly. 

It is found that the changes of the intra- and inter- band $B(E2)$ values are not large compared to the prediction of the $B(E2)$ reduction for $^{21}_{\ \Lambda}$Ne. This corresponds to the small deformation changes by a $\Lambda$ hyperon in $^{25}_{\ \Lambda}$Mg. 

%
%


\newpage 
\bibliography{submit} 

\end{document}